# Low frequency seismic responses and the challenge for acquisition

*Mark A. Meier\*, The University of Houston*

**Summary**

Low frequency seismic responses have considerably different characteristics than conventional band responses and require acquisition technologies that are capable of meeting far greater requirements. Seismic sources must deliver forces at lower frequencies that are considerably larger than the forces delivered by modern sources at conventional band frequencies in order to achieve comparable signal-to-noise ratios for many traditional interface-related seismic responses. Source efforts that are only comparable to conventional band source efforts are not adequate. Low frequency seismic responses from certain non-interface related impedance changes may be greater, but still require improved low frequency seismic sources.

**Introduction**

Seismic survey design procedures typically include evaluation of a subsurface target and prediction of the seismic response, then determination of equipment and survey characteristics required to achieve stakeholder objectives. The procedure is typically well-informed by experience with prior seismic surveys and existing equipment and commercial technologies. However, when pursuing new seismic response aspects never previously measured or observed, there is no prior experience to reckon upon. A return to fundamentals is helpful in designing experiments, developing foundational empiricisms, and establishing first generation specifications.

Interface-based models have been prevalent throughout the history of reflection seismology. The terminology implies seismic responses composed of reflections from a series of interfaces. My examination of seismic responses at low frequencies is a comparative evaluation referencing conventional frequency band concepts with which there is much experience and understanding. It begins with consideration of a single interface in the context of basic information on how far-field seismic radiation and ambient earth noise scale with frequency. Next, the single layer, defined by two bounding interfaces, is examined. I then depart from interface-based modelling to consider the Wolf ramp. The response and noise characteristics in the low frequencies are very different and imply different requirements for a low frequency source effort.

**Theory**

Comparison of responses across frequency bands requires stipulation of which of the seismic wave's particular physical properties are to be observed. This is because the manner in which pressure or particle velocity varies with frequency is different from how particle displacement or particle acceleration varies with frequency. I will reference the pressure or particle velocity, both behaving similarly across frequencies, to describe amplitudes of seismic waves.

Two fundamental factors that greatly affect low frequency seismic acquisition are the characteristics of far-field seismic radiation and ambient noise. In the latter case, it is well established that ambient noise levels generally increase with lower frequencies, particularly in the octaves near one hertz. Studies of ambient noise, in both land and marine environments, consistently show increases in pressure or particle velocity amplitudes at rates between 12 and 30 dB per octave (Berger, 2004; Gabrielson, 1995; Norris and Johnson, 2007; Urick, 1984). The analyses in this abstract assume 18 dB per octave. This might be considered a reasonable, though minimum rate of increase likely to be encountered in most exploration regions given nearby field and production activity commonly present.

Far-field pressure and particle velocity seismic amplitudes depend on force delivered by the source and are proportional to frequency (Miller and Pursey, 1954; Meier, 2016). That is, if all factors unrelated to the source are ignored; i.e., attenuation and transmission losses, etc., the far-field amplitude at one frequency will be one half that of the frequency one octave higher if the source is delivering the same force at both frequencies. The effects of attenuation and transmission loss can vary extremely from region to region, so the approach here is to account only for effects related to the source. This is equivalent to assuming a medium having no attenuation, or an inverse-Q of zero. The results and conclusions of this study can be adjusted to account for losses as might be assumed for application in a particular medium.

The synthetic seismograms that follow assume an incident Ricker wavelet defined by the frequency of peak amplitude (peak frequency). The seismic response is backscattered signal from the model geology assuming an incident Ricker wavelet with a peak frequency of choice. The model is restricted to one dimension, and the response is for normal incidence or, correspondingly, zero offset. The simulated noise for a given synthetic trace has the same spectral frequency as the incident Ricker wavelet under consideration. Noise amplitudes vary to represent the assumed 18 dB per octave increase with lower frequency.

The signal-to-noise ratio (S/N) and, in particular, how S/N changes with frequency is a very useful parameterization. A definition must always be provided. The root mean square

# Low frequency seismic responses and the challenge for acquisition

(rms) amplitude of a sinusoidal signal is the peak amplitude divided by square root of two. It would seem reasonable to consider a sinusoidal signal added to a random sequence with an equivalent rms amplitude as having a S/N equal to one. Therefore, the ratio between the peak sinusoidal amplitude divided by square root of two, and the rms amplitude of the random sequence is used to define S/N. By analogy, I will define a synthetic seismogram S/N as the ratio between the peak amplitude of the synthetic seismogram response divided by square root of two, and the rms amplitude of the simulated noise. So, a response signal having peak amplitude of 1 and noise with an rms amplitude of one divided by the square root of two is defined to be a S/N of 1, or 0 dB.

**Examples**

The following examples examine seismic responses from three different geological features; a single interface, a single layer, and a Wolf ramp. The responses assume a seismic source capable of delivering the same peak force at all frequencies. Relative amplitudes between frequencies in a given seismic response are true, but cannot be compared across seismic responses of different geological features.

Interface Response
The normal incidence reflection from a single interface has an amplitude that depends on the impedance contrast across the interface. The reflection amplitude is frequency independent. However, the far-field amplitude from a source delivering equal force at all frequencies falls 6 dB per octave for lower frequencies. This means the interface response grows weaker with lower frequencies if the peak force delivered by the source remains the same for any given Ricker frequency distribution. The interface response is shown in Figure 1a for Ricker wavelets with peak frequencies from 1 to 4 hertz. The interface is located at a two way travel time of two seconds. Each trace represents the response for an incident Ricker wavelet with a peak frequency increasing in geometric progression with trace number from left to right. This produces a linear scale in terms of octave count. The time scale includes one second on either side of the interface, and is considerably compressed relative to conventional seismic plotting in order to accommodate convenient viewing of the much lower frequency content.

The peak amplitude of the interface response at 4 hertz is four time greater than the peak amplitude at 1 hertz. If the same response amplitude is desired at lower frequencies, then the seismic source force output must increase by 6 dB, or doubling, per octave. This requirement is in stark contrast to a displacement limited seismic vibrator whose nominal force output falls by 12 dB per octave with lower frequency. The force fall-off for marine sources is typically even greater (Hegna and Parks, 2011). In real seismic data, the effect of propagation losses through the seismic medium reduces the fall in response amplitude with lower frequency somewhat since attenuation reduces higher frequency amplitudes relative to lower frequencies. A very simple attenuation model could be applied to reduce the response amplitude by 6 dB per octave with higher frequency, thereby flattening the response amplitude with frequency. Realistic models are likely to be more complicated depending on the frequency dependence of Q.

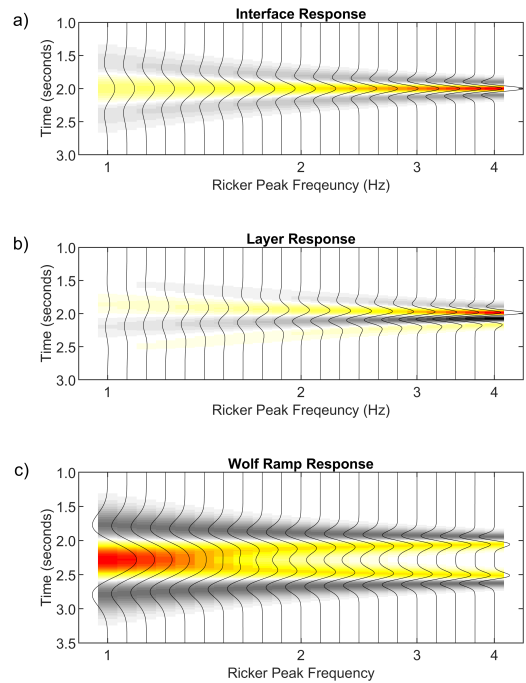

Figure 1: The seismic response from three geological features; a) an interface, b) a layer, and c) a Wolf ramp.

Layer Response
A layer consists of two interfaces bounding top and bottom. Under the weak scattering assumption, or otherwise assuming the successful removal of internal multiples and correction for the top interface transmission coefficient, the response may be modeled as a linear combination of two interface responses weighted by their corresponding reflection coefficients. The layer response considered here is for a layer in an otherwise homogeneous medium, meaning the reflection coefficient from the bottom interface is equal and opposite to that of the top interface. The normal incidence reflection from each interface, individually, is frequency independent and has an amplitude that depends on impedance contrast between the layer and outside medium. However, the interface reflections overlap one another and

# Low frequency seismic responses and the challenge for acquisition

destructively interfere at the lower frequencies. Consequently, the layer response is frequency dependent, and grows weaker with lower frequency due to the combined effects of destructive interference and 6 dB per octave loss in far-field amplitude with lower frequencies. The layer response grows weaker more rapidly with lower frequency than the single interface response.

Figure 1b shows the layer response for Ricker wavelets with peak frequencies from 1 to 4 hertz. The top interface of the layer is located at 2.0 seconds, and the bottom interface is located 62.5 milliseconds later. Full resolution of this layer, based on the effective duration criterion (Meier and Lee, 2009), requires a Ricker wavelet with a peak frequency of 9.7 hertz. Maximum tuning amplitude occurs with a 6.2 hertz Ricker wavelet (Kallweit and Wood, 1982). From 4 to 1 hertz, the effects of destructive interference are rapidly increasing and diminishing the response amplitude.

Wolf Ramp Response
A medium in which there are no impedance discontinuities, i.e., no interfaces, will still produce a backscatter response if the impedance is changing. An example is a medium which transitions linearly with depth from one velocity to another over some depth interval while density remains constant. The velocities outside the depth interval are constant with depth, and equal the end-ramp velocities. The analytical solution to normal incidence backscatter from this model was developed by Wolf (1936) with some recent descriptions by Liner and Bodmann (2010). The reflection amplitude and phase vary with frequency. Frequencies that are sufficiently high have negligible reflection amplitude whereas frequencies that are sufficiently low have a reflection amplitude that is mostly dependent on the contrast of impedances above and below the ramp depth interval.

Figure 1c shows a Wolf ramp response with Ricker wavelets from 1 to 4 hertz. The velocity above the ramp is 2000 meters per second and the velocity below the ramp is 3000 meters per second. The ramp starts at 2.0 seconds and linearly transitions from 2000 to 3000 meters per second over a depth interval of 700 meters. In spite of the loss in far-field amplitude of 6 dB per octave, the response amplitude increases substantially with lower frequencies. Beyond the frequencies shown, the response amplitude grows increasingly weak above 4 hertz. At increasingly lower frequencies below 1 hertz, it eventually behaves similarly to a single interface response.

The Wolf ramp response illustrates the lower frequency bands can "see" different aspects of the earth's impedance structure than the conventional band. Regions of slower impedance transitions that produce ignorable backscatter responses at conventional band frequencies may produce substantial backscatter responses in the lower frequency bands. These structures may dominate the lower frequency responses over customary structures readily identified in conventional band responses. For imaging, the concept of a smooth velocity model, that is regions of slowly varying impedance where there is seismic wave transmission, only, and no reflection or backscattering, may need considerable adjustment when considering lower frequencies.

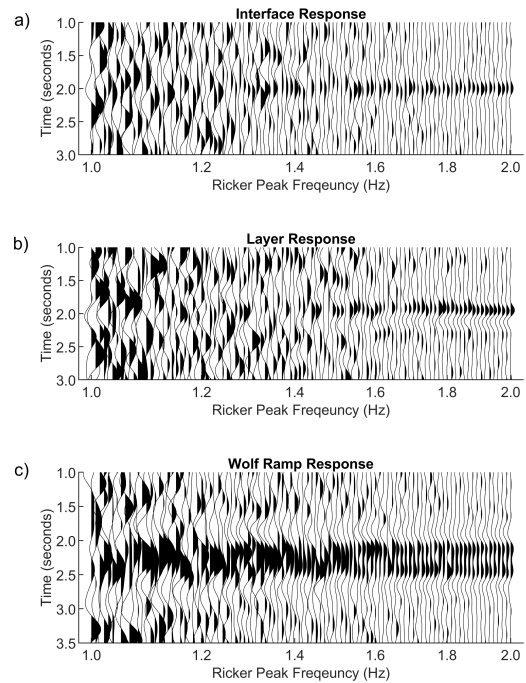

Figure 2: Seismic responses with noise increasing at 18 dB per octave with lower frequencies.

Noise Model
The relative strength or weakness of a seismic response means very little in the absence of noise. Without noise, any gain may be applied to the response constrained only by numerical limits imposed by bit architecture. The noise model used here is generated by a random variable and has the same spectral content as the Ricker wavelet used in generating the synthetic seismogram responses. The noise rms amplitude increases with lower frequencies by 18 dB per octave.

Figure 2 shows the interface, layer, and Wolf ramp responses with the noise model added. Because of the high rate of increased noise with lower frequencies, only a single octave of the response is shown; from 1 to 2 hertz. As in Figure 1, each trace represents the response for an incident Ricker wavelet with a peak frequency increasing in geometric progression with trace number from left to right; however,

# Low frequency seismic responses and the challenge for acquisition

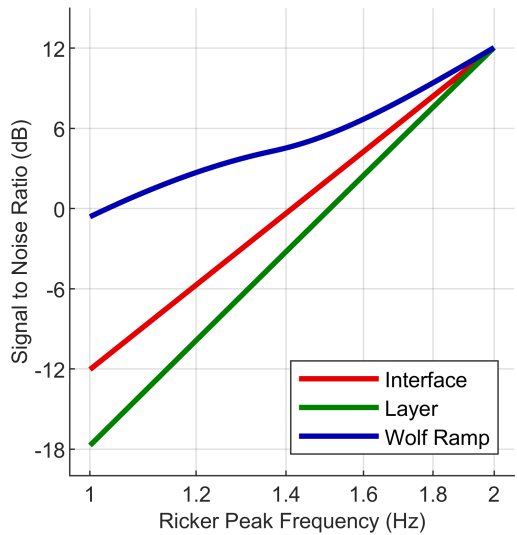

Figure 3: The signal to noise ratio is shown for the octave from 1 to 2 hertz.

there are many more traces and the wiggle trace display parameters are adjusted to more closely resemble a standard seismic display. The time scale is considerably compressed relative to conventional seismic plotting in order to accommodate convenient viewing of the much lower frequency content. The noise amplitude is adjusted so that the S/N is 12 dB on the 2 hertz trace. This is done so that the response at 2 hertz is clearly visible. Though, the ability to deliver 12 dB S/N at 2 hertz remains beyond current seismic practice, it serves as useful reference here to illustrate the challenge of delivering an additional low frequency octave when the noise is increasing at 18 dB per octave.

The interface and layer responses grow weaker with lower frequencies, as was shown in Figure 1, while the noise grows stronger. The Wolf ramp response grows stronger with lower frequencies but is overwhelmed by the more rapidly growing noise. Though it may be somewhat subjective to claim a particular frequency where the response becomes lost in the noise for each case, it is clear that the layer response is lost at a higher frequency, and the Wolf ramp response is lost at a lower frequency.

A plot of S/N for the interface, layer, and Wolf ramp responses are shown in Figure 3. Reduction in S/N of 24 dB, 30 dB, and 12 dB are observed for each of the responses, respectively, for the octave going from 2 to 1 hertz. Recall that the response amplitudes were created assuming a constant peak ground force available at all frequencies. However, seismic sources that can deliver conventional band forces at frequencies down to one hertz are not available. Even if they were, there is still a very substantial shortcoming to achieving comparable S/N. In this octave, it is necessary to increase seismic responses by 12 to 30 dB to achieve comparable S/N. The requirement may be somewhat alleviated by including attenuation effects in the model, and may reduce the S/N differential by about 6 dB per octave depending on the attenuation model and its dependence on frequency. However, this still leaves 6 to 24 dB of additional response amplitude that must come from increased source effort at 1 hertz over 2 hertz. This is in addition to the source effort required to achieve S/N of 12 dB at 2 hertz.

Increased source effort comes most efficiently from increased force levels. This may be in the form of seismic sources that can deliver larger peak forces, or larger arrays of seismic sources. The S/N increases proportionally with increased force. Another means comes from longer source duration. This may be in the form of longer sweep times or larger numbers of sweeps. However this is much less efficient, improving S/N only by the square root of increased duration. So, improving S/N by 30 dB may come from delivering a force about 30 times greater, or from source durations about 1000 times greater, or from some combination of the two.

**Conclusions**

Modelling seismic responses and noise demonstrates that source efforts must be dramatically improved to achieve S/N at lower octaves that are comparable to those in the conventional seismic frequency band. Achieving a given S/N at 1 hertz requires a source effort approximately an order of magnitude greater than that at 2 hertz. This is in addition to the source effort required to achieve S/N of 12 dB at 2 hertz. This likely corresponds to source efforts that are approximately two orders of magnitude greater than those in the conventional band. A ten-fold increase in source effort may be achieved by increasing peak force capability by an order of magnitude or by increasing source durations by two orders of magnitude, or some combination of the two. It is not reasonable to pursue low frequency responses with source efforts comparable or smaller than that used in conventional band acquisition. The substantially greater requirements for low frequency acquisition source effort suggests new and radically different technologies are necessary that can deliver greater peak forces for greater durations in both land and marine environments.

**Acknowledgements**

The author is grateful to the University of Houston for its support, and dedication of effort and resources towards establishment of the Low Frequency Seismic Technologies Consortium.

# Low frequency seismic responses and the challenge for acquisition